\documentclass[aps,pra,floatfix,twocolumn,superscriptaddress]{revtex4-1}
\usepackage{amsmath,amssymb}
\usepackage{graphicx}
\usepackage{epsfig}
\usepackage{ulem}
\usepackage[usenames]{color}

\begin{document}

\title{Interplay between collective modes\\ in hybrid electron gas--superconductor structures}



\author{M.~V.~Boev}
\affiliation{A.V.~Rzhanov Institute of Semiconductor Physics, Siberian Branch of Russian Academy of Sciences, Novosibirsk 630090, Russia}
\affiliation{Novosibirsk State Technical University, Novosibirsk 630073, Russia}

\author{I.~G.~Savenko}
\affiliation{Center for Theoretical Physics of Complex Systems, Institute for Basic Science (IBS), Daejeon 34126, Korea}
\affiliation{Basic Science Program, Korea University of Science and Technology (UST), Daejeon 34113, Korea}
\affiliation{A.V.~Rzhanov Institute of Semiconductor Physics, Siberian Branch of Russian Academy of Sciences, Novosibirsk 630090, Russia}

\author{V.~M.~Kovalev}
\affiliation{Novosibirsk State Technical University, Novosibirsk 630073, Russia}

\date{\today}

\begin{abstract}
We study hybridization of collective plasmon and 
Carlson-Goldman-Artemenko-Volkov modes in a hybrid system, consisting of a two-dimensional layers of electron gas in the normal state and superconductor, coupled by long-range Coulomb forces.
The interaction between these collective modes is not possible in a regular single-layer two-dimensional system since they exist in non-overlapping domains of  dimensionless parameter $\omega\tau$, where $\omega$ is the external electromagnetic field frequency and $\tau$ is electron scattering time. Thus, in a single-layer structure, these modes are mutually exclusive.
However, the coupling may become possible in a hybrid system consisting of two separated in space materials with different properties, in particular, the electron scattering time.
We investigate the electromagnetic power absorption by the hybrid system and reveal the conditions necessary for the hybridization of collective modes. 
\end{abstract}

\maketitle


\section{Introduction}
Mesoscopic systems in normal (not superconducting) state lodge elementary excitations, such as excitons and plasmons.
They determine the response of the material to external perturbations.
Plasmons represent collective modes of particle density fluctuations and play an essential role in the material response to external alternating electromagnetic (EM) fields. Their dispersion law and damping strongly depend on the dimensionality of the system.
Indeed, in the case of three-dimensional (3D) electron plasma, the plasmon branch is with a good accuracy dispersionless, whereas two-dimensional (2D) plasmons are characterized by a square-root dispersion relation and vanish in long-wavelength limit (at least, in the framework of the quasistatic approximation).
Many-component 2D systems like an electron-hole 2D gas are characterized by two plasmon branches with linear (acoustic plasmon) and square-root (optical plasmon) dispersions.

If with a decrease of temperature the system undertakes a phase transition to another state like a Bose-Einstein or superconducting (SC) condensate, the ground state changes and new elementary excitations describing low-energy properties of the system arise.
This can lead to drastic modification of the material response to external EM fields, which becomes especially important in view of resent progress in plasmonics, optoelectronics, and high-temperature condensation phenomena.

In the pioneering work of Bardeen Cooper and Schriffer (BCS)~\cite{BCS}, only single-particle excitations characterized by the SC gap were studied.
Later, Anderson~\cite{Anderson}, Bogoliubov~\cite{Bogoliubov} and Vaks, Galitskii, and Larkin~\cite{VaksGalitskiiLarkin} developed a more extensive theory, which includes the collective excitations of Cooper pairs.
They showed, that in a neutral Fermi system with an attractive interaction between fermions, collective modes possess a soundlike dispersion.
A charged Fermi system, instead, demonstrates the collective modes with the frequency pushed towards the plasmon frequency of the 3D charged electron gas.
This frequency is usually much larger than the BSC gap and consequently it does not play role in conventional phenomena in SCs.

These modes correspond to oscillations of the SC order parameter.
More precisely, the Anderson-Bogoliubov mode represents the oscillation of the phase of the order parameter, whereas the Higgs mode corresponds to the oscillations of its amplitude.
These collective modes play crucial role in the gauge-invariant response of superconductors to external EM perturbations~\cite{Arseev}.

Later on, Carlson and Goldman~\cite{CG1, CG2, CG3} experimentally discovered another type of collective modes, which occur in the vicinity of the transition temperature $T_c-T\ll T_c$ in superconductors.
This experimental finding stimulated theoretical studies.
In the dirty-sample case, there was suggested the model of Schmid and Sch\"on~\cite{SS}, whereas in the clean (low-disorder) limit, there was developed the model by Artemenko and Volkov (AV)~\cite{AV1, AV2}.
Qualitatively, AV mode corresponds to the out-of-phase oscillations of normal and SC currents in such a way that it prevents the emergence of net charges in the system.
The frequency of this mode is smaller than the plasmon frequency of normal 3D electrons, thus suppressing the interaction between them.


In the meanwhile, the coupling between single-particle and collective modes in hybrid systems consisting of a 2D electron layer and a 2D condensate of Bose particles represent actively
developing areas of research recently~\cite{RefAlt, Imamoglu, KovChap}.
In particular, there emerge such phenomena as the magnetoplasmon Fano resonances in uniform magnetic fields~\cite{BKS} and amplification of the incident light~\cite{OurPRBKus, OurPRLKus}.
The progress in theoretical proposals and the lack of well-established experimental platforms motivates the study of the interplay of different collective modes in hybrid systems.
In this article, we investigate the hybridization of collective plasmon and AV modes in a hybrid semiconductor-superconductor structure (where one of the subsystems is in the SC state). 
As we have already pointed out, the plasmon mode has a gapless dispersion in 2D case.
We will show that it opens a possibility of interaction between AV modes and plasmons.

At first sight, despite the gaplessness of the plasmon mode, the interaction between them is not possible since the AV mode exists in the domain $\omega\tau\ll1$ (here $\tau$ is electron-impurity scattering time), whereas the plasmon mode exist at $\omega\tau\gg1$, and these two conditions do not overlap.
However, their interaction can become possible if the normal electron gas layer (having the plasmon mode) and the SC layer (hosting the AV mode) are separated in space and electrons in each layer have different scattering times.
Thus, if we denote as $\tau_n$ the scattering time of electrons in the normal layer and $\tau_s$ the one for quasiparticles in the SC layer, the simultaneous existence of AV and plasmon modes is possible when $\tau_s^{-1}>\omega>\tau_n^{-1}$. This is a necessary criterion for the coexistence of both the modes in one sample. However, it is not a sufficient condition for their hybridization. What are the other conditions? This is the main question, which we address in this article.


%
%
%
\begin{figure}[!t]
\includegraphics[width=0.44\textwidth]{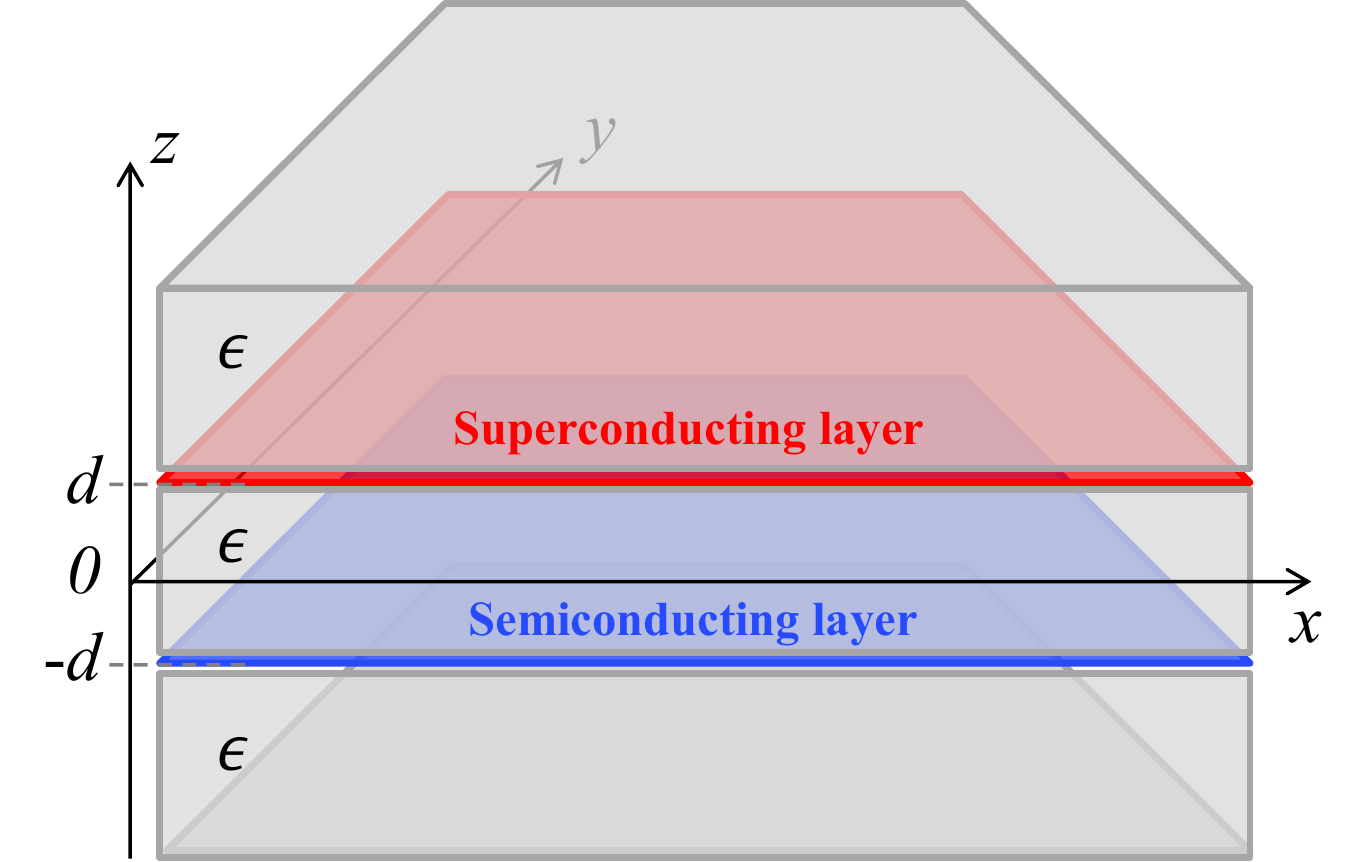}
\caption{System schematic. 
A hybrid structure consisting of a superconducting layer located at $z=d$ and a semiconductor layer at $z=-d$. The layers are thus separated by the distance $2d$ and surrounded by the environment with the effective dielectric constant $\epsilon$.}
\label{System}
\end{figure}

\section{Longitudinal dielectric function and the dispersion relation}

Let us consider a 2D layer of a semiconductor containing normal electrons with the equilibrium density $N_n$ in the vicinity of a 2D SC layer containing both normal single-electron excitations above the SC gap $\Delta(T)$, where $T$ is temperature \textcolor{black}{(we will use $\Delta/T\ll1$)}, and SC Cooper pairs described by the order parameter (Fig.~\ref{System}). 
The total electron density in SC layer is $N_s$, whereas the the density of SC electrons we denote as $n_s$.

Furthermore, we expose the system to an external EM field, which represents a plane wave with 3D wave vector $\textbf{Q}=(\textbf{k},Q_z)$, where the 2D wave vector $\textbf{k}=(Q_x,Q_y)$ is its in-plane projection.
The field creates electric currents in both the layers,
\begin{gather}\label{Eq1}
\textbf{j}_s=\sigma_s(\textbf{E}_0+\textbf{E}_s^i),\\\nonumber
\textbf{j}_n=\sigma_n(\textbf{E}_0+\textbf{E}_n^i),
\end{gather}
where $\textbf{E}_0$ is the in-plane component of external EM field, which we assume to be equal in both layers (valid as long as $Q_zd\ll1$);
$\textbf{E}_{n,s}^i$ are the  internal EM fields, induced in the corresponding layers due to the oscillations of the particle densities;
 \begin{gather}\label{Eq7}
     \sigma_n=\frac{e^2N_n\tau_n/m_n}{1-i\omega\tau_n}
 \end{gather}
is a conventional dynamical Drude conductivity of the degenerate electron gas in the semiconductor, with $m_n$ being the electron effective mass; $\sigma_s$ describes the current response of single-particle excitations in the SC layer.
It should be noted, that in Eq.~\eqref{Eq1} we omitted the factor $\exp(i\textbf{kr}-i\omega t)$ since all the time and in-plane position-dependent quantities are proportional to this factor.

The induced fields depend on $z$, $\textbf{E}_{n,s}^i\equiv\textbf{E}^i(z=\pm d)$, where $\textbf{E}^{i}$ can be expressed via the scalar potential in the framework of quasistatic approximation, $\textbf{E}^{i}=-\nabla \varphi^{i}$.
The scalar potential, in turn, satisfies the Poisson equation,
\begin{gather}\label{Eq2}
\left(\frac{d^2}{dz^2}-k^2\right)\varphi^{i}(z)=-\frac{4\pi}{\epsilon} [\delta \rho_s\delta(z-d)+\delta \rho_n\delta(z+d)],
\end{gather}
where $\delta \rho_{n,s}$ are deviations of charge densities from their equilibrium values, and $\epsilon$ is a dielectric function of the material.
Combining~\eqref{Eq1} and~\eqref{Eq2} with the continuity equation, we find
 \begin{gather}\label{Eq3}
    \left(
        {1+\frac{2\pi ik}{\epsilon\omega}\sigma_s \atop \frac{2\pi ik}{\epsilon\omega}e^{-2kd}\sigma_n} \quad
        {\frac{2\pi ik}{\epsilon\omega}e^{-2kd}\sigma_s \atop 1+\frac{2\pi ik}{\epsilon\omega}\sigma_n}
    \right)
    \left(
        {\delta \rho_s \atop \delta \rho_n}
    \right) = \frac{kE_0}{\omega}
    \left(
        {\sigma_s \atop \sigma_n}
    \right)
 \end{gather}
and
 \begin{gather}\label{Eq4}
    \left(
        {j_s \atop j_n}
    \right) =
    \left(
        {\left[1 + \frac{2\pi ik}{\epsilon\omega}\left(1-e^{-2kd}\right)\sigma_n\right]\sigma_s \atop
         \left[1 + \frac{2\pi ik}{\epsilon\omega}\left(1-e^{-2kd}\right)\sigma_s\right]\sigma_n}
    \right) \frac{E_0}{\varepsilon(k,\omega)},
 \end{gather}
where we have introduced the longitudinal dielectric function,
 \begin{gather}\label{Eq5}
    \varepsilon(k,\omega) =
    \left(1 + \frac{2\pi ik}{\epsilon\omega}\sigma_s\right)\left(1 + \frac{2\pi ik}{\epsilon\omega}\sigma_n\right) +\\\nonumber
    +
    \left(\frac{2\pi ke^{-2kd}}{\epsilon\omega}\right)^2\sigma_s\sigma_n.
 \end{gather}
Here and in what follows we assume that $\textbf{k}$ and $\textbf{E}_0$ only have $x$ in-plane components.

Eq.~\eqref{Eq5} is similar in form to the standard one, describing generic two-component systems.
Equation $\varepsilon(k,\omega)=0$ determines the dispersion and damping of collective modes.
The key quantity here is $\sigma_s$, which has to be found accounting for the interaction of single-particle excitations in SC layer with the Cooper pair condensate.

Let us, first, switch off the interaction between the layers and analyze the collective modes in each layer separately.
Formally, the noninteracting case corresponds to putting $d\rightarrow\infty$ in Eq.~\eqref{Eq5}.
Then $\varepsilon(k,\omega)=0$  splits into two independent conditions,
 \begin{gather}\label{Eq6}
       1 + \frac{2\pi ik}{\epsilon\omega}\sigma_{n}=0;~~~
       1 + \frac{2\pi ik}{\epsilon\omega}\sigma_{s}=0.
 \end{gather}
Let us consider these cases separately.

\subsection{Plasmons in the layer containing 2DEG}

%
%
%
%
Substituting Eq.~\eqref{Eq7} in~\eqref{Eq6}, we find the plasmon dispersion and its damping due to the electron-impurity scattering,
 \begin{gather}\label{Eq8}
     \omega=\omega_n\sqrt{1-\frac{1}{(2\omega_n\tau_n)^2}}-\frac{i}{2\tau_n};\,\,\,\,
     \omega_n=\sqrt{\frac{2\pi e^2kN_n}{\epsilon m_n}}.
 \end{gather}
The plasmon exists if $2\omega_n\tau_n>1$.
In the case of weak scattering, when $\omega_n\tau_n\gg1$, the plasmon damping is much smaller in comparison with its frequency, and in the EM power absorption spectrum, the plasmons are seen as well-resolved resonances.
The spatial dispersion of Drude conductivity~\eqref{Eq7} does not play role in~\eqref{Eq8} and can be neglected if $\omega\gg kv_n$, where $v_n$ is the Fermi velocity of electrons in the semiconducting layer.


\subsection{Artemenko-Volkov modes in the SC layer}

To find the dispersion law of the AV mode, we have to, first, find the conductivity $\sigma_s$ of single-particle excitations above the SC gap.
In their original work~\cite{AV1}, Artemenko and Volkov used the quasi-classical approach based on the kinetic equations.
Later, the validity of their results was confirmed by the quantum field theory methods~\cite{Ovchinnikov}.
We will thus follow the simpler original Boltzmann equation approach, where the AV modes are found by analyzing the longitudinal dielectric function~\eqref{Eq6}.
The calculation of $\sigma_s$ for a 3D superconductor can be found elsewhere~\cite{LozovikApenko}, and we adopt it for 2D SC layer just presenting here the result \textcolor{black}{(see Appendix for the details of the derivation)},
 \begin{gather}
 \label{Eq9}
     \sigma_s =
    \sigma_{0s}\cfrac{\omega^2-u^2k^2+i\omega\eta_s+ik^2v_s^2\omega\tau_sJ_\omega}
    {\omega^2-u^2k^2+ik^2v_s^2\omega\tau_sJ_\omega},
 \end{gather}
where
 \begin{gather}\label{Eq10}
    \sigma_{0s} = \frac{e^2N_s\tau_s}{m_s},\,\,\,u=v_s\sqrt{\frac{7\zeta(3)}{2\pi^3}\frac{\Delta}{T}},\,\,\,\eta_s=\cfrac{n_s}{N_s}\cfrac{1}{\tau_s},\\\nonumber
    J_\omega=\cfrac{\ln(1/\omega\tau_s)}{\pi}.
 \end{gather}
Here we introduced a static Drude conductivity of normal electrons in SC layer, $\sigma_{0s}$;
$v_s$ is the Fermi velocity in SC layer;
$u$ is the phase velocity of AV mode (see the description below), $\zeta(3)$ is the Riemann zeta-function.
We note also that $u\ll v_s$.

Formula~\eqref{Eq9} is valid under the conditions $\omega,~kv_s,~\tau_s^{-1}\ll\Delta\ll T$ and $(\tau_skv_s)^2\ll\omega\tau_s\ll1$~\cite{LozovikApenko}.
Using these assumptions and combining Eqs.~\eqref{Eq9} and~\eqref{Eq6}, we find
 \begin{gather}\label{Eq11}
    1+i\frac{\omega^2_s\tau_s}{\omega}\left(1+i\frac{\omega\eta_s}{\omega^2-u^2k^2+ik^2v_s^2\omega\tau_sJ_\omega}\right)=0,\\
    \nonumber
    \textrm{where}~~\omega_s=\sqrt{\frac{2\pi e^2kN_s}{\epsilon m_s}}.
 \end{gather}
In the limiting case $\omega^2_s\tau_s\gg\omega$~\cite{LozovikApenko}, there exists an analytical solution,
 \begin{gather}\label{Eq12}
\omega^2=u^2k^2-ik^2v_s^2\omega\tau_sJ_\omega-i\omega\eta_s.
\end{gather}
We note again, that this solution describes a weakly-damped sound-like AV mode in the frequency range
 \begin{gather}\label{Eq13}
\left(\frac{\Delta}{T}\right)^2\ll\omega\tau_s\ll\frac{\Delta}{T}\ll1.
\end{gather}

\begin{figure}[!t]
\includegraphics[width=0.48\textwidth]{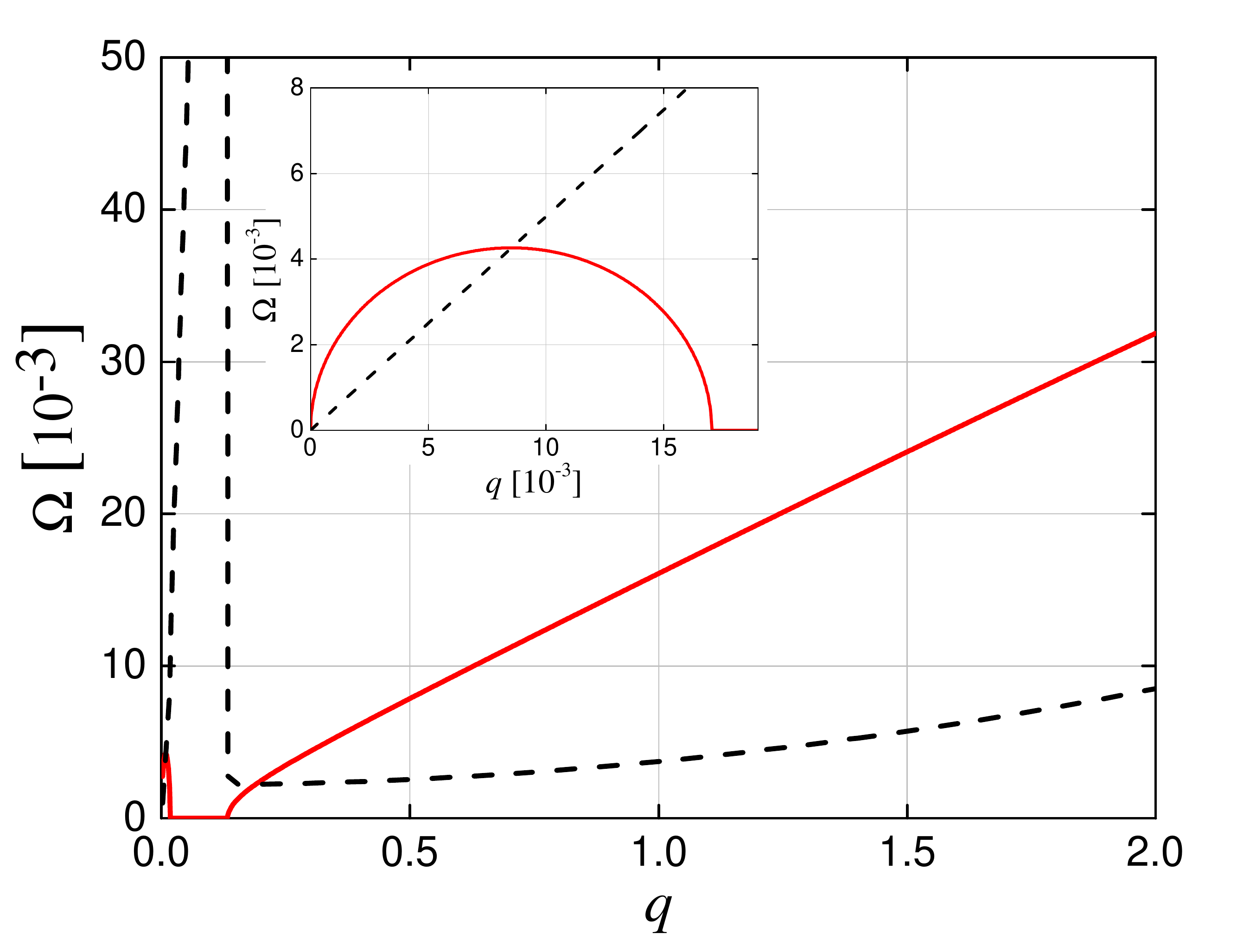}
\caption{
Dimensionless complex-valued  eigen frequency $\Omega=\omega\tau_s$ as a function of dimensionless wave vector $q=(\omega_s\tau_s)^2$ for isolated SC layer. 
\textcolor{black}{We use
$a_B/(v_s\tau_s)=1/10$, $\Delta/T=1/10$.}
Red curve stands for the dispersion $\textmd{Re}\,\Omega$ whereas the dotted black curve shows the absolute value of the damping $|\textmd{Im}\,\Omega|$. 
Inset shows the zoom-in of the low-wave vector domain, where the qAB mode is visible and not suppressed by the damping.}
\label{Fig1SCmodes}
\end{figure}

The numerical solution of Eq.~\eqref{Eq11} gives two collective branches in the isolated 2D SC layer.
The first dispersion is presented in Fig.~\ref{Fig1SCmodes}, where we use dimensionless frequency $\Omega=\omega\tau_s$ and wave vector $q=(\omega_s\tau_s)^2$. 
\textcolor{black}{Also, we use $m_s=m_n$, $a_{B}=\epsilon/e^2m_s$ in Fig.~\ref{Fig1SCmodes} and all the figures which follow.}
We see that the AV mode characterized by the linear dispersion $uk$ lies above the almost horizontal part of red-dotted line, which stands for the lower bound $(\Delta/T)^2$ in Eq.~\eqref{Eq13}.

The second mode is located in the low-$q$ domain, as shown in Inset in Fig.~\ref{Fig1SCmodes}. 
At $q\ll1$, this mode is weakly-damped due to the smallness of its imaginary part ($\sim q$) in comparison with its eigen frequency $\sim \sqrt{q}$. 
The dispersion of this mode can be found analytically from Eq.~\eqref{Eq11}. 
Indeed, in the limit $k\rightarrow 0$, Eq.~\eqref{Eq11} reduces to
 \begin{gather}\label{Eq13.1}
    1+i\frac{\omega^2_s\tau_s}{\omega}\left(1+i\frac{\eta_s}{\omega}\right)=0,
\end{gather}
which gives the dispersion of the 2D plasmon,
\begin{eqnarray}\label{Eq13.2}
 \omega&=&\omega_s\sqrt{\frac{n_s}{N_s}}\sqrt{1-\frac{N_s}{n_s}\left(\frac{\omega_s\tau_s}{2}\right)^2}-i\frac{\omega_s^2\tau_s}{2},
\end{eqnarray}
existing when $n_s/N_s>(\omega_s\tau_s)^2/4$.
On condition $n_s/N_s\gg(\omega_s\tau_s)^2/4$, formula~\eqref{Eq13.2} 
turns into not damped 2D plasmon oscillations of SC electrons, 
\begin{gather}
    \label{Eq13.3}
    \omega \approx \omega_s\sqrt{\frac{n_s}{N_s}}=\sqrt{\frac{2\pi e^2kn_s}{\epsilon m_s}}.
\end{gather}

It should be noted, that in contrast with the normal-state systems, where the plasmon exists if $\omega_n\tau_n\gg1$, the mode Eq.~\eqref{Eq13.3} exists at $\omega_s\tau_s\ll1$. 
This mode reminds the Anderson-Bogoliubov mode~\cite{Anderson, Bogoliubov} studied at $T=0$ in 3D superconductors. 
It represents plasmon oscillations of SC electrons, which density at $T=0$ equals to the total electron density in the superconductor. 
The difference is that in our 2D case, at $T\neq0$ this mode is determined by the density of SC electrons $n_s\neq N_s$. 
We will call it quasi-Anderson-Bogoliubov mode (qAB). The qAB oscillations of the SC condensate at $T\neq0$ are accompanied by the appearance of induced charges in the system which, obviously, interact with normal electrons.
They exist in a superconductor at nonzero temperatures and play the role of a friction force influencing the oscillations of the condensate. 
At $T_c-T\ll T_c$ (the case which we consider in this article), the density of normal electrons in the SC layer is of the order of total electron density $N_s$.
It explains why the imaginary part in Eq.~\eqref{Eq13.2} is proportional to $\omega_s^2\propto N_s$.

%
%
%
\begin{figure}[!b]
\includegraphics[width=0.48\textwidth]{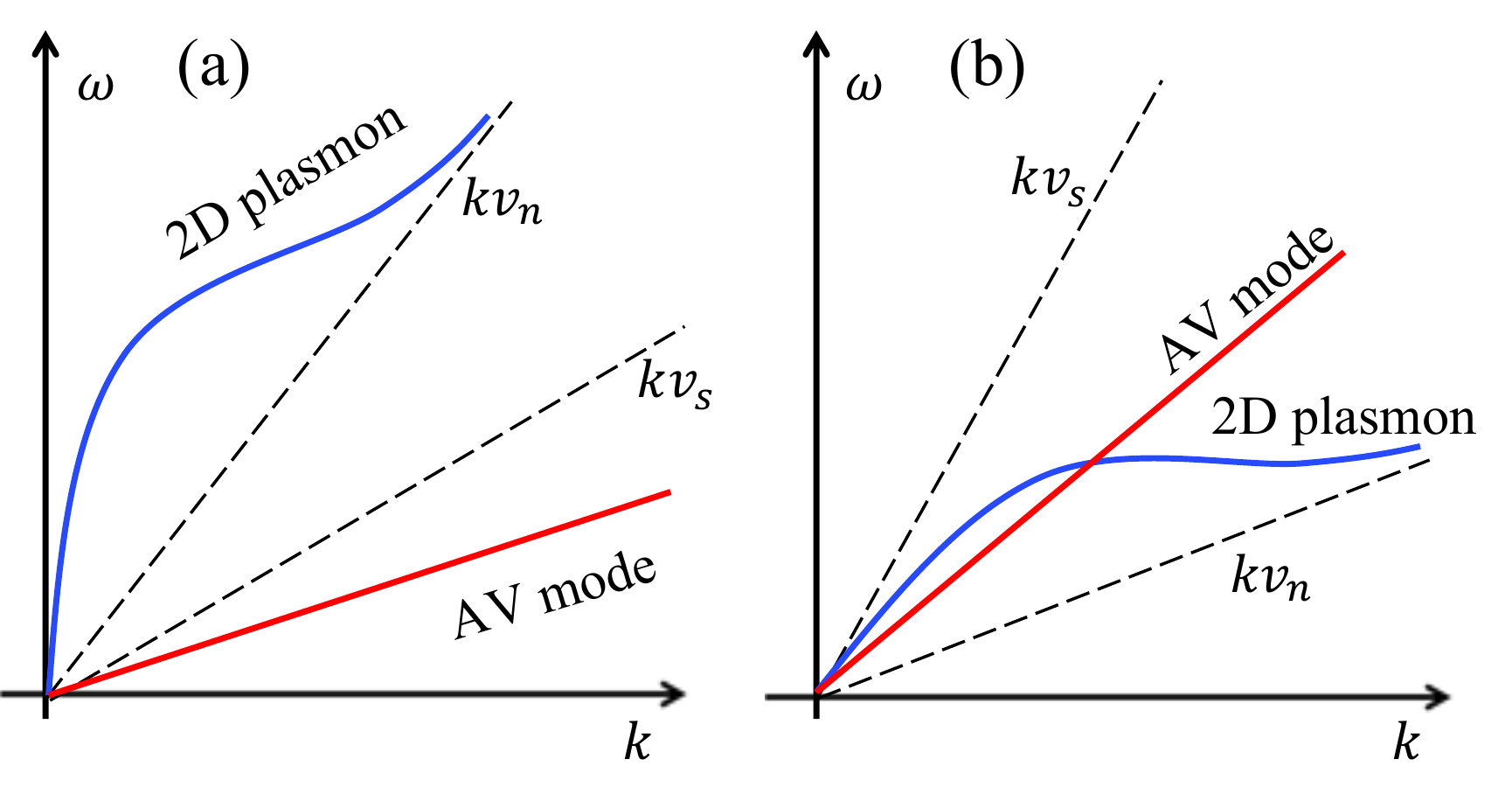}
\caption{Qualitative behavior of non-interacting plasmon and AV modes in the case of a) $v_n>v_s$ and b) $v_s>v_n$. 
(The qAB mode is not shown.)
}
\label{Dispersions}
\end{figure}
%
%
%


\section{Interaction between the modes}

We have already found, that the plasmon mode of SC layer lies above the boundary of the particle-hole continuum, $\omega_n\gg kv_n$, whereas the AV mode has the phase velocity $u\ll v_s$.
Hence there exist two limiting cases, depending on the ratio between the Fermi velocities in SC, $v_s$, and normal 2DEG, $v_n$, layers (Fig.~\ref{Dispersions}).
Let us consider both of these scenarios independently.


\subsection{Modes interaction if $v_n>v_s$}

In this case, the dispersions of the modes do not intersect. 
Thus at a given value of the wave vector $k$, 
these modes are excited independently in the system at different frequencies of external EM field.
Nevertheless, the presence of the other layer results in the inter-layer Coulomb interaction of the particles and produces an additional damping of the collective mode which is excited.
Let us analyse this damping.

In the frequency range when the plasmon mode exists, $\omega\sim\omega_n$, $\omega_n\tau_n\gg1$ we can simplify  formulas for the conductivities~\eqref{Eq7} and~\eqref{Eq9},
\begin{gather}\label{Eq14}
\sigma_n\approx\frac{e^2N_n}{-i\omega m_n},\,\,\,\sigma_s\approx\sigma_{s0}=\frac{e^2N_s\tau_s}{m_s},
\end{gather}
and the dispersion equation $\varepsilon(k,\omega)=0$ reduces to
\begin{gather}\label{Eq15}
\left(1-\frac{\omega_n^2}{\omega^2}\right)\left(1+i\frac{\omega_s^2\tau_s}{\omega}\right)=-i\omega_s\tau_s\frac{\omega_n\omega_s}{\omega^2}e^{-4kd}.
\end{gather}
Its iterative solution gives the renormalization of the plasmon dispersion and an additional damping,
\begin{gather}\label{Eq16}
    \omega=\omega_n
        \left
         (1-\omega_s\tau_s
          \frac{\omega_s\tau_s(1+e^{-4kd})+2i\sqrt{N_n/N_s}} {4N_n/N_s+\omega_s^2\tau_s^2(1+e^{-4kd})^2}e^{-4kd}
        \right).
\end{gather}
We see that if $kd$ is not very large and $N_n/N_s\ll\omega_s^2\tau_s^2$, the presence of SC layer can drastically change the plasmon dispersion resulting in a dramatic \textit{red shift} of the plasmon frequency.

If, instead, the frequency lies in the range of the AV mode, $\omega\approx uk$, the Drude conductivity of electrons in the semiconducting layer can be replaced by its static limit $\sigma_n\approx e^2_n\tau_n/m_n$, and the dispersion equation reads
%
\begin{eqnarray}
\label{Eq17}
&&\left(1+i\frac{\omega_n^2\tau_n}{\omega}\right)
\left[1+i\frac{\omega^2_s\tau_s}{\omega}\left(1\right.\right.
\\
\nonumber
&&\left.\left.
+i\frac{\omega\eta_s}{\omega^2-u^2k^2+ik^2v_s^2\omega\tau_sJ_\omega}\right)\right]=
\left(\frac{\omega_n^2\tau_n}{\omega}\right)
\left(\frac{\omega^2_s\tau_s}{\omega}\right)
\\
\nonumber
&&~~~~~~~~~~~~~~
\times\left(1+i\frac{\omega\eta_s}{\omega^2-u^2k^2+ik^2v_s^2\omega\tau_sJ_\omega}\right)e^{-4kd}.
\end{eqnarray}
The bare AV mode can exist under the condition $\omega^2_s\tau_s\gg\omega$, as we have discussed above.
Due to the relation $\omega_n\tau_n\gg\omega_s\tau_s$, we can assume $\omega\ll \omega^2_s\tau_s\ll \omega^2_n\tau_n$ in Eq. \eqref{Eq17}, and we find that the AV mode is not renormalized in the presence of semiconducting layer (in the first-order perturbation theory with respect to the parameter $\omega^2_s\tau_s/\omega$).
One can find the next-order corrections with respect to   $1\ll \omega^2_s\tau_s/\omega\ll \omega^2_n\tau_n/\omega$;
however, they are small and we will not discuss them further.

\subsection{Modes interaction if $v_n<v_s$}

\begin{figure}[!t]
\includegraphics[width=0.45\textwidth]{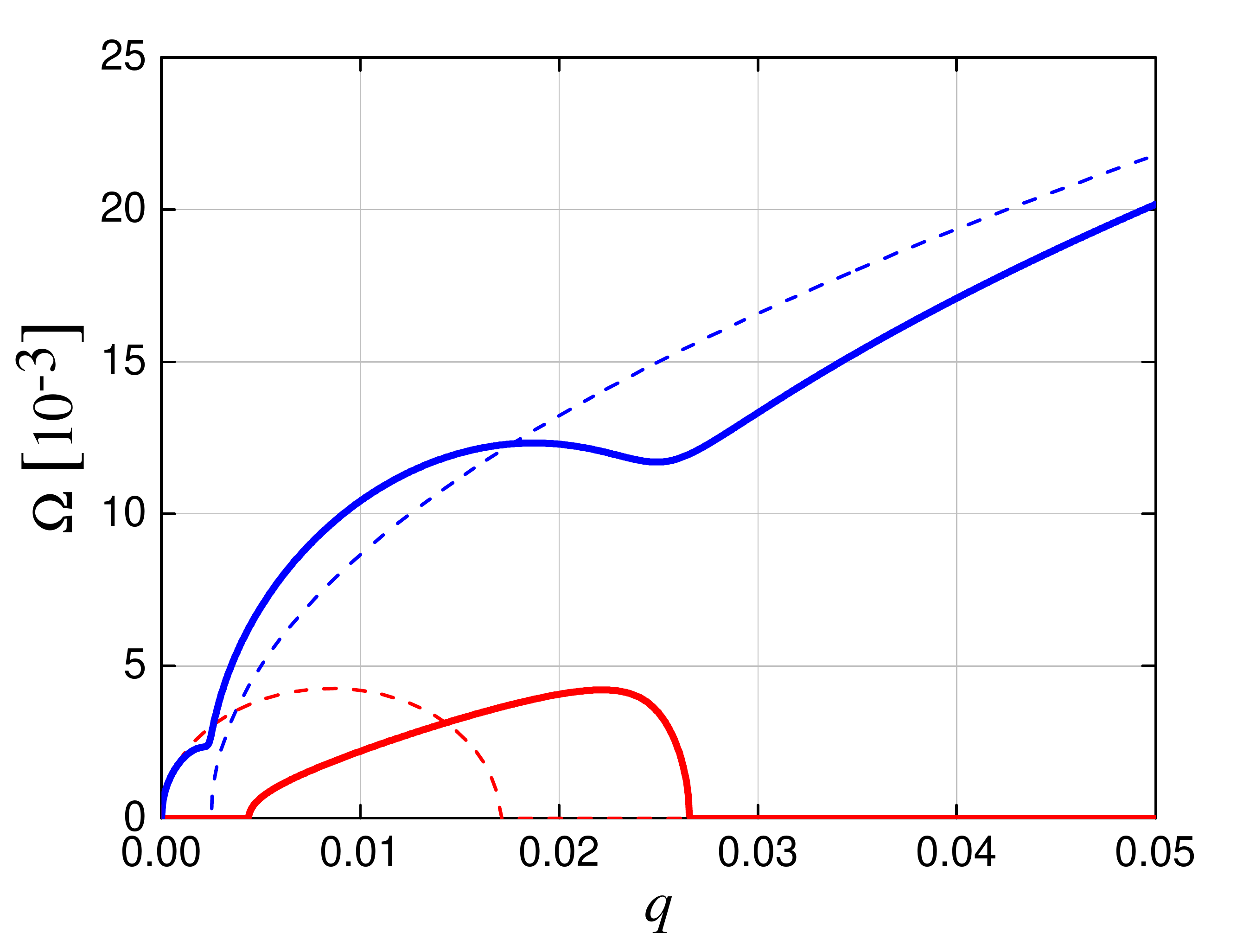}
\caption{
Dispersions of hybrid modes in the system (solid curves).
The dispersions of bare qAB and plasmon modes are shown in dashed red and blue curves, respectively.
\textcolor{black}{
We use $a_B/v_s\tau_s=1/10$, $\Delta/T=1/10$, $\tau_n/\tau_s=100$, $N_s/N_n=100$; and for solid curves, $d/v_s\tau_s=100$.}}
\label{figuries2and3}
\end{figure}

\begin{figure}[!t]
\includegraphics[width=0.45\textwidth]{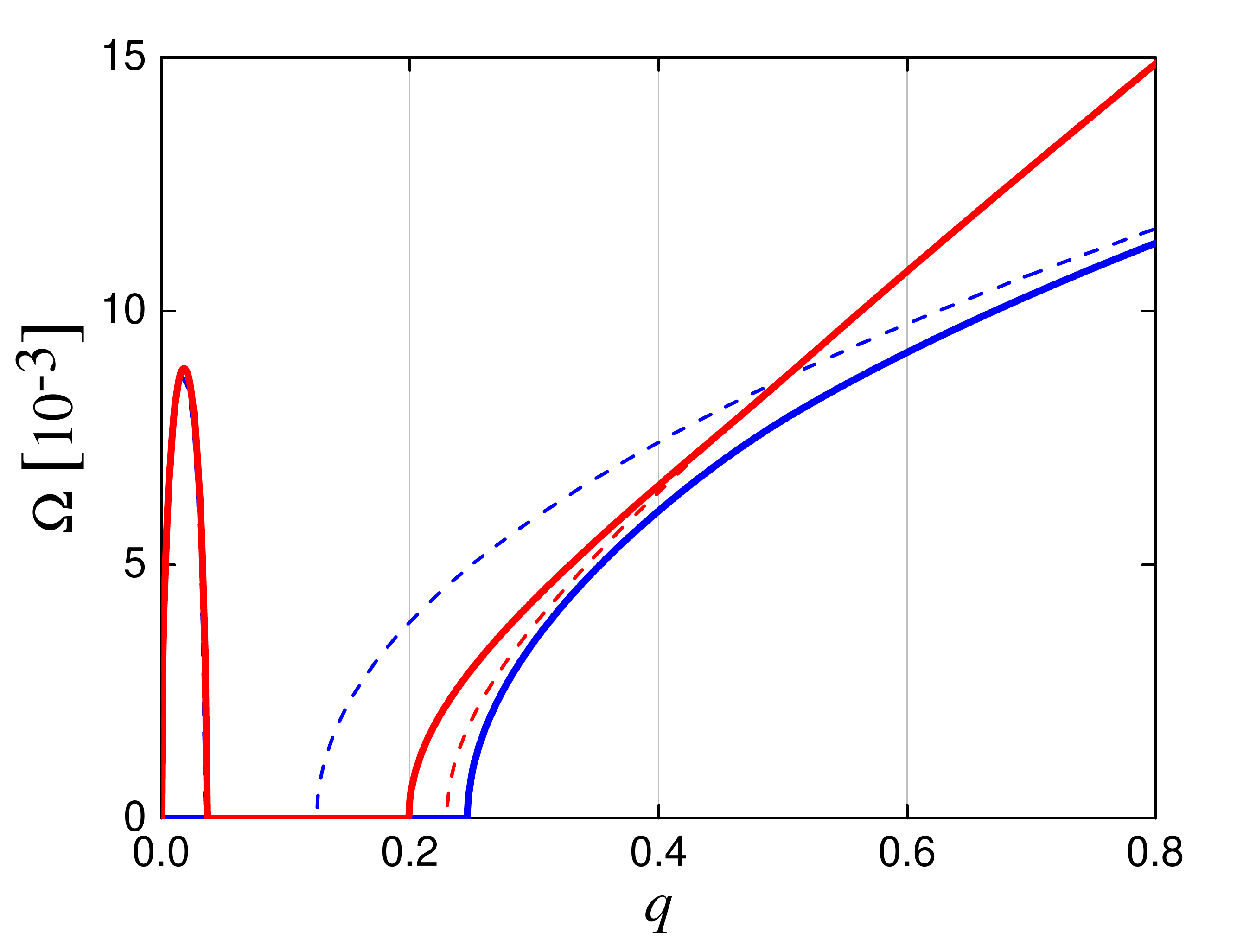}
\caption{
Dispersions of hybrid modes in the system (solid curves).
The dispersions of bare AV and plasmon modes are shown in dashed red and blue curves, respectively.
\textcolor{black}{
We use $a_B/v_s\tau_s=1/10$, $\Delta/T=1/7$, $\tau_n/\tau_s=100$, $N_s/N_n=5000$; and for solid curves, $d/v_s\tau_s=10$.}
}
\label{figuries4and5}
\end{figure}

In this case, the density of electrons in the semiconducting layer is smaller than the total density of electrons in the SC layer, $N_n<N_s$. 
The intersection of the plasmon mode (of semiconducting layer) with both qAB and AV modes (of SC layer) is possible, depending on the system parameters, as it is shown by dashed curves in Fig.~\ref{figuries2and3} and Fig.~\ref{figuries4and5}.
%
%
%

We analyze the interaction between the modes numerically (see solid curves presented in Fig.~\ref{figuries2and3} and Fig.~\ref{figuries4and5}). 
The main conclusion is that the plasmon mode of semiconducting layer interacts with qAB mode stronger than with AV mode since in the latter case, the modes repel each other shifting their positions in the $q$-axis. 
This situation can be explained if we recall that the qAB mode corresponds the oscillation of SC elections and it is accompanied by the oscillations of induced electric fields. 
In contrast, the AV mode corresponds to the anti-phase oscillations of SC and normal electrons in SC layer resulting in the quasi-neutrality. 
Therefore, since the interaction between the layers has Coulomb nature, it is not surprising that the plasmon mode interacts with qAB mode much stronger.


\section{Electromagnetic field power absorption}

Let us expose the system to an external EM field and study light-matter interaction. 
We will do a numerical analysis of EM power absorption $\textmd{Re}\,(j_nE_0^*)$ [where $j_n$ is defined in Eq.~\eqref{Eq4}] in the case when $v_s>v_n$ and, thus $N_n<N_s$. 
In the opposite regime, the interaction of the modes is weak. 

\begin{figure}[!t]
\includegraphics[width=0.45\textwidth]{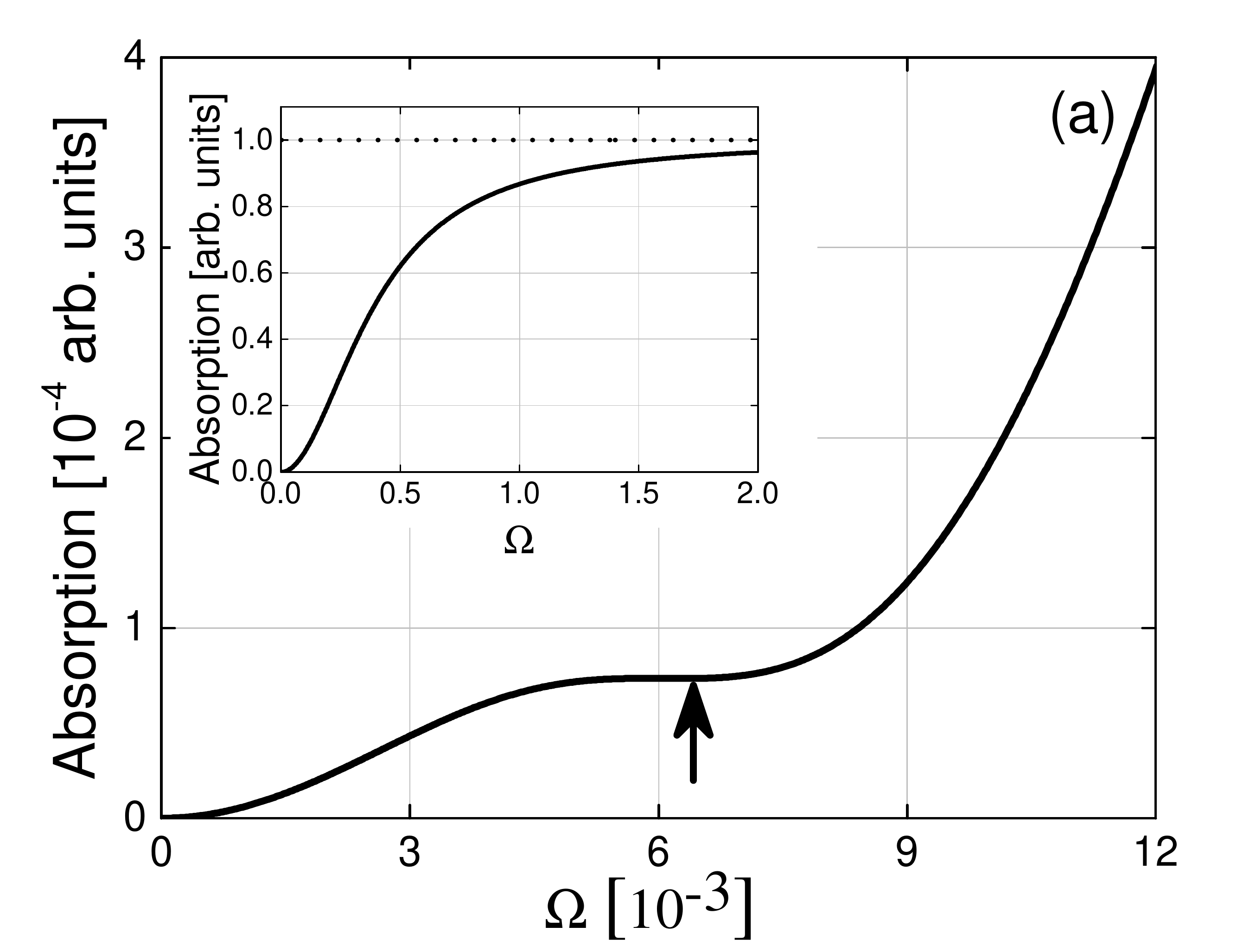}
\includegraphics[width=0.45\textwidth]{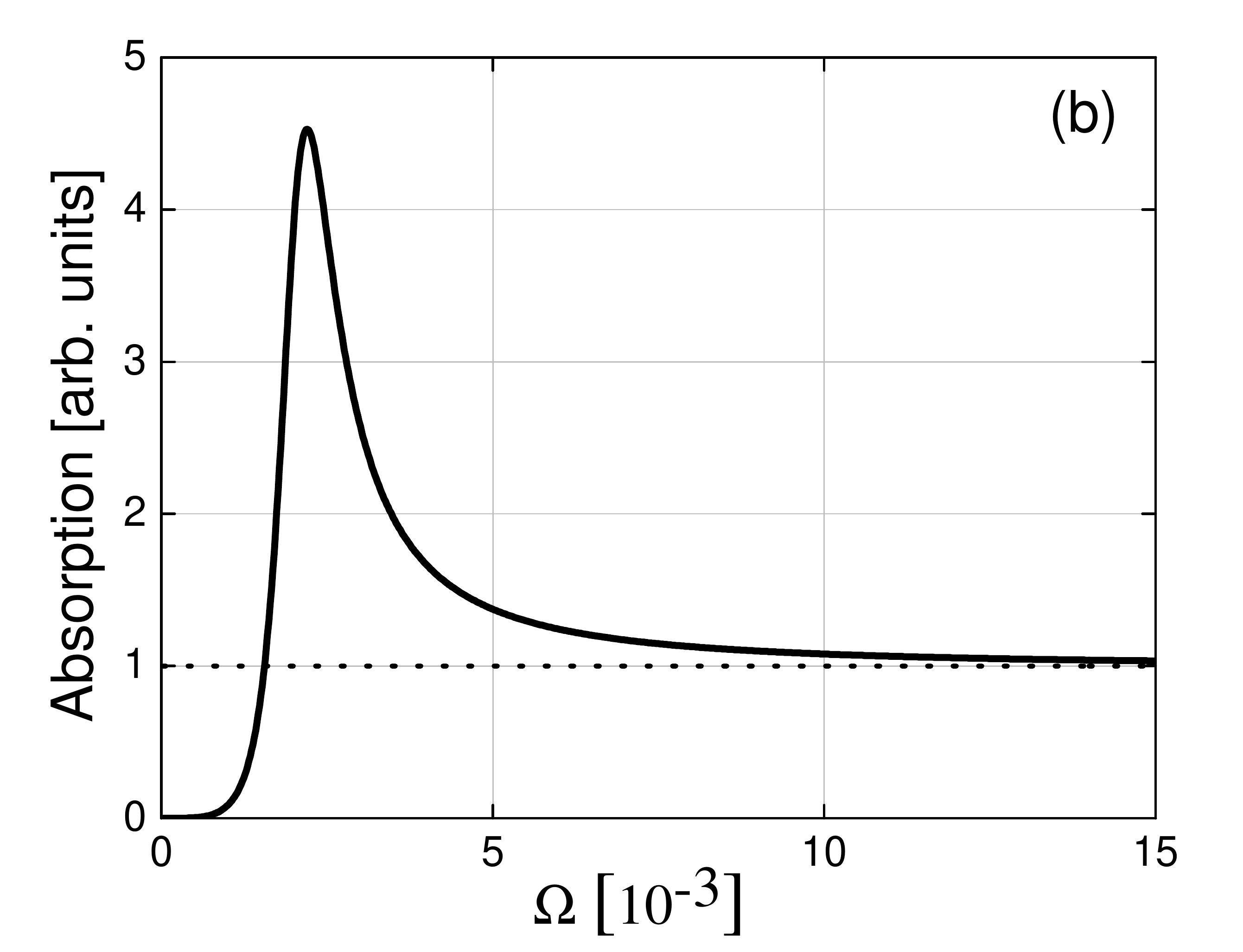}
\caption{EM power absorption by the isolated SC layer as a function of dimensionless frequency $\Omega$ given in arbitrary units. 
(a) EM power absorption by the AV mode. 
\textcolor{black}{
We use $q=0.4$, $a_B/v_s\tau_s=1/10$, $\Delta/T=1/7$, $\tau_n/\tau_s=100$, $N_s/N_n=5000$.}
Inset shows the absorption spectrum at large values of $\Omega$; (b) EM power absorption by the qAB mode.
\textcolor{black}{
We use $q=0.001$, $a_B/v_s\tau_s=1/10$, $\Delta/T=1/10$, $\tau_n/\tau_s=100$, $N_s/N_n=100$.}
}
\label{figuries6and7}
\end{figure}

First, we switch off the interaction between the layers. 
The EM power absorption of an isolated semiconducting layer is characterized by a standard Lorenz-shaped resonance at the plasmon frequency. 
A more interesting situation arises in the isolated SC layer due to the presence of both the qAB and AV modes. The AV mode is quasi-neutral, and hence it interacts with the external EM field weakly, thus the EM power absorption is negligibly small.
Indeed, the approximate analytical solution Eq.~\eqref{Eq12} if substituted into Eq.~\eqref{Eq9} gives $\sigma_s=0$ and $j_n=0$ reflecting the quasi-neutral nature of AV mode and nearly zero EM power absorption. 
The exact numerical calculation of the power absorption in the absence of interaction between SC and semiconducting layers ($d\rightarrow\infty$), presented in Fig.~\ref{figuries6and7}(a), supports this conclusion. 
At large frequencies $\omega\gg ku$, the absorption in Fig.~\ref{figuries6and7}(a)
reaches a plateau with the value equal to the static Drude absorption of EM radiation of normal electrons in SC layer, as it is seen from Eq.~\eqref{Eq9}. 
Indeed, at $\omega\gg ku$ the conductivity of SC layer resembles the static Drude conductivity $\sigma_s\approx\sigma_{s0}$. 
The arrow in Fig.~\ref{figuries6and7}(a) indicates EM power absorption at the position of AV mode which is negligibly small in comparison with the power absorption corresponding to the plateau value. 

In contrast, the qAB mode strongly interacts with external EM field and demonstrates a pronounced peak of absorption with the amplitude much greater than the plateau value, as it is shown in Fig.~\ref{figuries6and7}(b).  

\begin{figure}[!t]
\includegraphics[width=0.45\textwidth]{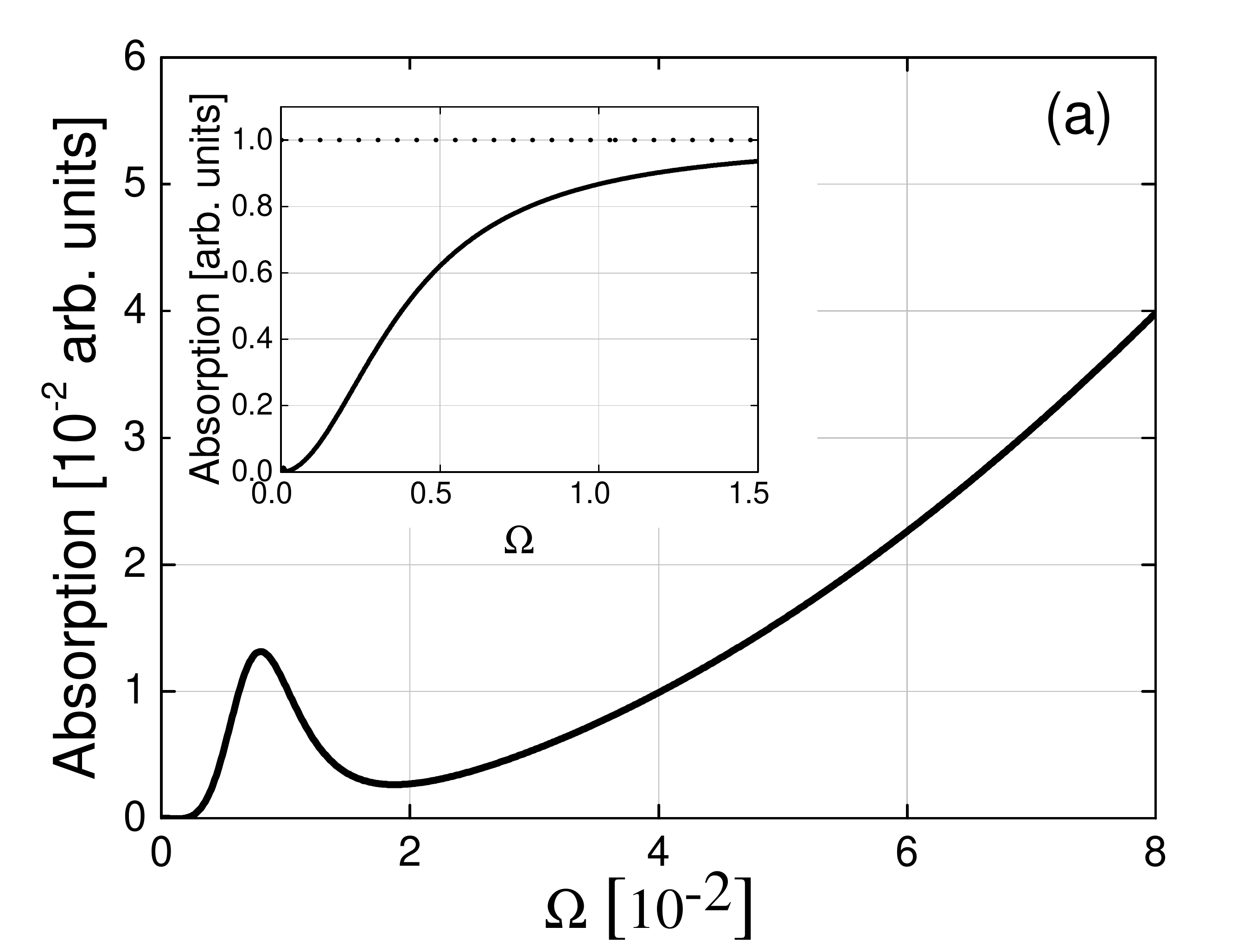}
\includegraphics[width=0.45\textwidth]{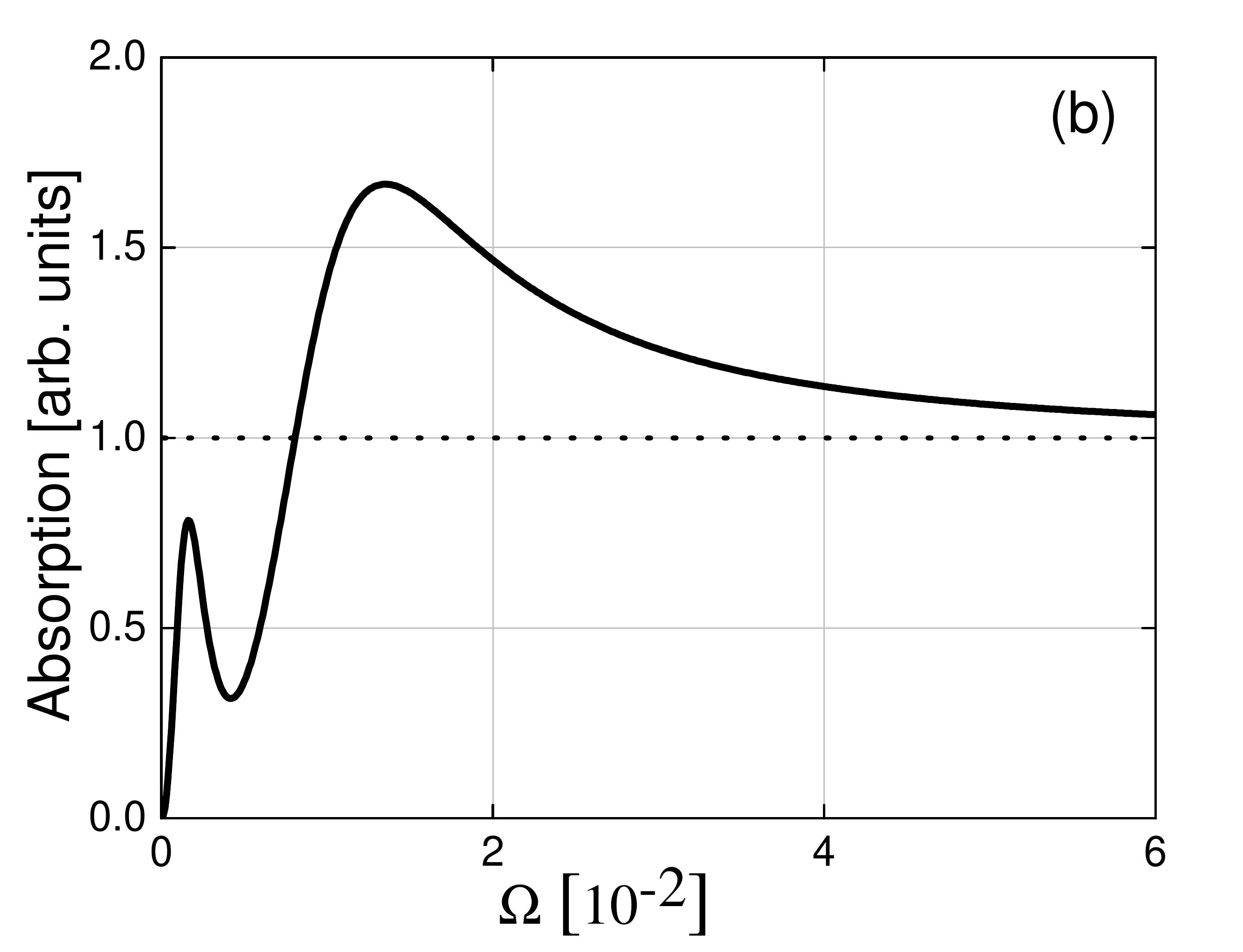}
\caption{
EM power absorption by the hybrid system as a function of dimensionless frequency $\Omega$ given in arbitrary units. 
(a) EM power absorption by the AV-plasmon mode. 
\textcolor{black}{
We use $q=0.4$, $a_B/v_s\tau_s=1/10$, $\Delta/T=1/7$, $\tau_n/\tau_s=100$, $N_s/N_n=5000$, $d/v_s\tau_s=10$.}
Inset shows the absorption spectrum at large values of $\Omega$; 
(b) EM power absorption by the qAB-plasmon mode.
\textcolor{black}{
We use $q=0.006$, $a_B/v_s\tau_s=1/10$, $\Delta/T=1/10$, $\tau_n/\tau_s=100$, $N_s/N_n=100$, $d/v_s\tau_s=100$.}}
\label{figuries8and9}
\end{figure}

Furthermore, let us switch the interlayer interaction on and consider the EM power absorption by the hybrid system when the collective modes of SC are hybridised with the plasmon mode in the semiconducting layer (Fig.~\ref{figuries8and9}). 
Figure~\ref{figuries8and9}(a) shows the EM power absorption by the hybrid AV-plasmon modes. 
Weak hybridization of the modes results in two new eigen modes. 
One of them almost resembles the original plasmon mode, and the other mostly possesses the properties of bare AV mode. 
From our previous discussion of the light absorption by the bare modes, we expect that the quasi-plasmon mode (the hybrid mode which mostly inherited the properties of the plasmon mode) will give a pronounced resonance in EM absorption spectrum, whereas the other, quasi-AV mode (the hybrid mode which mostly inherited the properties of the AV mode) will not be seen. This reasoning is supported by the exact numerical calculations, shown in Fig.~\ref{figuries8and9}(a).  
We see a single resonance at the frequency of the quasi-plasmon mode with a shifted position with respect to bare plasmon dispersion and a  pronounced Lorenz shape. 
The quasi-AV mode is not seen in the figure. 

At large $\Omega$, the EM power absorption spectrum again reaches the plateau with the value equal to the Drude conductivity of the SC layer. 
The amplitude of the Lorenz-shaped resonance is smaller than the value at the plateau since the total density of electrons in the semiconducting layer is smaller than the density of electrons in SC layer, $N_n<N_s$.

Finally, EM power absorption by the hybrid qAB-plasmon mode demonstrates a double-resonance structure, indicating that both the new modes formed from the original qAB and plasmon modes (before the coupling) are extremely sensitive to external EM radiation.
We want to point out, that such shape of the spectrum can be attributed to the Fano resonance, which is expected in hybrid Bose-Fermi systems~\cite{BKS}.
And again, at large frequencies, the energy absorption reaches the plateau with the value equal to the Drude conductivity absorption of SC layer.


\textcolor{black}{
\section{Actual parameters to conduct the experiment and limitations}
Let us discuss the actual parameters of the sample to be used in the experiments. 
We will consider qAB mode first.
The Drude model is applicable if $k_Fl\gg 1$, where $l$ is the mean free path of normal electrons. 
We choose $N_s=10^{14}$~cm$^{-2}$ and $m_s=0.5$~$m_0$, which are typical for superconductors based on transition-metal dichalcogenides~\cite{japans}. 
Then $k_Fl\approx 1.38\times 10^{15} \tau_s$, 
and thus we can take $\tau_s=10^{-14}$~s. 
Note, with these parameters the resistivity amounts to kOhm (which corresponds to experimentally measured values),
}
\textcolor{black}{
while the Bohr radius becomes $a_B\approx \epsilon\cdot10^{-10}$~m. 
For $\epsilon=6$, the dimensionless ratio $a_B/v_F\tau_s$ used in all the plots is equal to $1/10$, that corresponds to the interlayer distance $d=100v_F\tau_s=1000a_B=600$~nm. 
In the normal layer, we use typical parameters $N_n=10^{12}$~cm$^{-2}$ and $\tau_n=10^{-12}$~s. 
They give for the ratio of static conductivities $\sigma_{s0}/\sigma_n(\omega=0)=1$.}

\textcolor{black}{
Second, let us consider AV mode. To bring the plasmon and AV mode dispersions together, we need a change of parameters. 
In our calculations, we make the ratio $N_s/N_n$ 50 times bigger. If we keep $N_n=10^{12}$~cm$^{-2}$ 
(the same as before), then the scattering times $\tau_s$ and $\tau_n$ should become $\sqrt{50}$ times smaller to keep the dimensionless ratios $\tau_s/\tau_n$ and $a_B/v_F\tau_s$ unchanged.  
Note, unlike qAB mode, AV mode starts feeling the presense of plasmons in the normal layer at an order of magnitude smaller interlayer distances, $d=10v_F \tau_s=100a_B=60$~nm.}

%
%
%
\textcolor{black}{
In the calculations and plots, we used dimensionless wave vector and frequency, $\Omega=\omega\tau_s\sim 0.01$. Then $\omega/2\pi\sim 0.1-1$~THz, which corresponds to the wavelength of the order of $0.1-1$~mm. 
For $q=0.4$ and $q=0.006$ we find $2\pi/k\sim 1~\mu$m and $\sim 70~\mu$m, respectively, which corresponds to the period of typical diffraction gratings (which are required in the experiments to excite the plasmons).}

\textcolor{black}{
Let us now discuss the limitations of our theoretical approach.
When a normal 2DEG is in the vicinity of a 2D superconductor, there might arise several phenomena, one of which is the proximity effect~\cite{RefHolm, RefProx2}, which might result in an induced superconductivity in the semiconducting layer.
In our calculations, we used a relatively large 2DEG-superconductor separation. Evidently, the smaller the separation, the more pronounced the coupling between the qAB, AV and plasmon modes. 
However, at very small separations, there will arise the proximity effect, which might quench the formation of collective modes and should be accounted for in engineering samples. 
It limits the separation distance to the coherence length of the particular superconductor. 
For instance, the coherence length amounts to several nanometers in the case of high-$T_c$ superconductors. 
}

\textcolor{black}{
An electron in 2DEG with energy below $\Delta$ can tunnel into superconductor due to the Andreev scattering~\cite{RefAndreev}.
However, if the distance between the layers is large and most of electrons are paired, the probability of such tunneling is small. 
If the superconductor is exposed to external light with the frequency above $\Delta$, the tunneling can be enhanced. Indeed, external EM fields excite electron and hole quasiparticles (which coexist with SC fluctuations in the superconductor). 
In this case, in addition to the Andreev scattering, there occurs common tunneling of quasiparticles into the superconductor.
Usually, this effect can also be disregarded when studying collective modes.
In particular, the frequency of external EM field which we consider in this article (to excite the modes) is smaller than $\Delta$.}


\section{Conclusion}
We have studied analytically and numerically the coupling of collective modes in a hybrid two-dimensional electron gas--superconductor structure in the vicinity of the critical temperature of superconducting transition of the superconductor. 
The layer containing the electron gas in normal state can be represented by a semiconductor or a metallic layer, used in hybrid NS contacts~\cite{RefAlt}. 
We have shown, that the superconducting layer lodges two collective modes with drastically different physical properties, which results in their different sensitivity to external electromagnetic fields. 
As a result, these modes couple differently with the plasmon mode of the normal electron gas spatially separated from the superconducting layer. 
It manifests itself in different spectra of the electromagnetic power absorption of the hybrid system. 






\acknowledgements
We have been supported by the Russian Foundation for Basic Research (Project No.~19-42-540011), the Ministry of Science and Higher Education of the Russian Federation (Project FSUN-2020-0004), and the Institute for Basic Science in Korea (Project No.~IBS-R024-D1).

\textcolor{black}{
\appendix*
\section{Derivation of the conductivity in 2D systems}
We present here the main steps of the derivation of conductivity~\eqref{Eq9} succeeding~\cite{LozovikApenko}. 
To find the form of SC conductivity (which describes the linear response of the system to external EM field, ${\bf j}_s=\sigma_s {\bf E}$), we can write the supercurrent within the Bogoliubov--de Gennes approach,
 \begin{gather}\label{Acur}
    {\bf j}_s = \frac{en_s{\bf p}_s}{m_s} + \frac{2e}{m_s}\sum\limits_{\bf p}{\bf p}f_{\bf p}({\bf p}_s,\phi),
 \end{gather}
where ${\bf p}_s = (\nabla\chi-2e{\bf A})/2 = m_s{\bf V}_s$, ${\bf V}_s$ is the velocity of SC flow, $\chi$ is the phase of the order parameter, $\phi = e\varphi + \partial_t\chi/2$, ${\bf A}$ and $\varphi$ are the vector and scalar potentials, respectively, and $f_{\bf p}({\bf p}_s,\phi)\equiv f_{\bf p}=n_{\bf p}({\bf p}_s,\phi)-n_{\bf p}^{(0)}$ is the nonequilibrium contribution to the distribution function of excitations in the superconductor. 
These excitations possess the following dispersion,
 \begin{gather}
    {\tilde\zeta}_{\bf p} = 
    \sqrt{{\tilde \xi}^2_{\bf p}+\Delta^2} + {\bf p}{\bf V}_s,~\textrm{with}
    \\
    {\tilde \xi}_{\bf p} = 
    \frac{{\bf p}^2}{2m} - \mu + \phi + \frac{{\bf p}^2_s}{2m},
 \end{gather} 
and their distribution function $n_{\bf p}({\bf p}_s,\phi)\equiv n_{\bf p}$ obeys the conventional kinetic equation
 \begin{gather}\label{kineq}
    \frac{\partial n_{\bf p}}{\partial t} + 
    \frac{\partial {\tilde\zeta}_{\bf p}}{\partial {\bf p}}\frac{\partial n_{\bf p}}{\partial {\bf r}} -
    \frac{\partial {\tilde\zeta}_{\bf p}}{\partial {\bf r}}\frac{\partial n_{\bf p}}{\partial {\bf p}} +
    I\{n_{\bf p}\} = 0,
 \end{gather}
where $I\{n_{\bf p}\}$ is the collision integral. 
Furthermore, we should linearize~\eqref{kineq} over ${\bf p}_s$ and $\phi$, and then find the deviation $f_{\bf p}$ and substitute it in~\eqref{Acur}. 
Then we can express ${\bf p}_s$ and $\phi$ via electric field strength ${\bf E}$. 
For that we need three equations (accounting for the fact that we deal with the vector ${\bf p}_s$, which has two components). 
Two of them result from the definitions of ${\bf p}_s$ and $\phi$,
 \begin{gather}\label{AE}
    \frac{\partial {\bf p}_s}{\partial t} - \nabla \phi = e{\bf E},
 \end{gather}
while the 
third one is the continuity equation,
 \begin{gather}\label{conteq}
    e\frac{\partial \delta n}{\partial t} + \nabla{\bf j}_s  = 0.
 \end{gather}
Here $\delta n$ is the deviation of electron density from equilibrium and it takes the following form by linearization,
 \begin{gather}\label{Adeln}
    \delta n = 
    2\sum\limits_{\bf p}
    \left\{
        f_{\bf p}\frac{\xi_{\bf p}}{\zeta_{\bf p}} + \phi
            \left[
                \left(\frac{\xi_{\bf p}}{\zeta_{\bf p}}\right)^2 \frac{\partial n^{(0)}_{\bf p}}{\partial \zeta_{\bf p}} -
                \frac{\Delta^2}{2\zeta^3_{\bf p}}\tanh\frac{\zeta_{\bf p}}{2T}
            \right]
    \right\},
 \end{gather}
where $\xi_\mathbf{p}=\tilde\xi_\mathbf{p}(\mathbf{p}_s=0,\varphi=0)$ and $\zeta_\mathbf{p}=\tilde\zeta_\mathbf{p}(\mathbf{p}_s=0,\varphi=0)$.
Solving the set of equations~\eqref{Acur}, \eqref{AE}, \eqref{conteq}, \eqref{Adeln}, and linearized~\eqref{kineq} we come up with formula~\eqref{Eq9} in the main text. 
}

\textcolor{black}{
Note, the work~\cite{LozovikApenko} treats 3D superconductors. 
In current 2D problem (which we deal with in this article), we suppose that the density of states of quasiparticles $G(\xi)$ does not include any singularities in the vicinity of the Fermi energy, like in work~\cite{LozovikApenko}. 
It means that performing usual transformations $\sum_{\bf p}\rightarrow G(\mu)\int d\xi$, we arrive at the same integrals over $\xi$ as in 3D case. 
Thus, the results are almost identical to ones reported in~\cite{LozovikApenko}. 
They coincide if we replace $3\rightarrow 2$ and take the electron density per unit area (instead of the electron density per unit volume).
}


\end{document}